\pdfoutput=1
\def\year{2021}\relax
\documentclass[letterpaper]{article}
\usepackage{aaai21}
\usepackage{times}
\usepackage{helvet}
\usepackage{courier}
\usepackage[hyphens]{url}
\usepackage{graphicx}
\usepackage{graphicx}
\usepackage{natbib}
\usepackage{caption}
\title{A SAT-based Resolution of Lam's Problem}

\author{Curtis Bright,\textsuperscript{\rm 1,2} Kevin K.~H.~Cheung,\textsuperscript{\rm 2} Brett Stevens,\textsuperscript{\rm 2} Ilias Kotsireas,\textsuperscript{\rm 3} Vijay Ganesh\textsuperscript{\rm 4}\\}

\affiliations{\textsuperscript{\rm 1}University of Windsor, \textsuperscript{\rm 2}Carleton University, \textsuperscript{\rm 3}Wilfrid Laurier University, \textsuperscript{\rm 4}University of Waterloo}

\usepackage{tikz}
\usepackage{multirow}
\usepackage{mathtools}
\usepackage[T1]{fontenc}
\usepackage{amsthm}
\newtheorem{theorem}{Theorem}

\begin{document}
\maketitle
\begin{abstract}
In 1989, computer searches by Lam, Thiel, and Swiercz experimentally
resolved Lam's problem from projective geometry---the
long-standing problem of determining if a projective plane of order
ten exists.  Both the original search and an independent
verification in 2011 discovered no such projective plane.
However, these searches were each performed
using highly specialized custom-written code
and did not produce nonexistence certificates.
In this paper, we resolve Lam's problem by translating
the problem into Boolean logic and use satisfiability
(SAT) solvers to produce nonexistence
certificates that can be verified by a third party.  Our work
uncovered consistency issues in both previous searches---%
highlighting the difficulty of relying on special-purpose search code
for nonexistence results.
\end{abstract}

\section{Introduction}

Projective geometry was developed in the 1600s by renaissance artists
and mathematicians in order to describe how to project a three dimensional
scene onto a two dimensional canvas.  Projective geometry has
the counter-intuitive property that \emph{any two lines
must meet}.  For example, a pair of train
tracks (parallel lines in three dimensions) when projected onto two dimensions
will meet on the horizon.

Despite an intensive amount of study for over 200 years some basic questions
about projective geometry remain unsettled.  For example---how many points
can a projective geometry have?  A geometry is said to be \emph{finite}
if it contains a finite number of points and a finite geometry
is said to have \emph{order} $n$ if every line contains $n+1$ points~\cite{dembowski}.

All finite projective geometries have been classified with the exception
of those having exactly two dimensions---the \emph{projective planes}.
The first order for which it is theoretically uncertain if a projective
plane exists is $n=10$.  Determining if such a projective plane exists has
become known as \emph{Lam's problem} after the work of~\cite{lam1989non}
experimentally showed that such a plane does not exist---work
that was later independently verified by~\cite{roy2011confirmation}.

As pointed out by~\cite{heule2017solving} there are currently
three kinds of solvers that are used to solve large but finite
decision problems like Lam's problem: special purpose solvers, constraint
satisfaction solvers, and satisfiability (SAT) solvers.  They note that
recently SAT solvers have become so strong that they are ``the best solution
in most cases''.  Even still, they note that some problems such as
Lam's problem have only been solved by special-purpose means:
\begin{quote}
An example where only a solution by [special purpose solvers]
is known is the determination that there is no projective plane of order 10\dots
\end{quote}
We remedy this situation by using a SAT solver to resolve
the most challenging subcase of Lam's problem.  Together with the recent SAT-based
results of~\cite{bright2020nonexistence,bright2020unsatisfiability} this provides
a complete resolution of Lam's problem with all
of the exhaustive search work completed by SAT solvers.

The previous searches done in Lam's problem remain fantastic achievements, but
a SAT-based resolution has two primary advantages.  First, it is more verifiable:
a third party can check the nonexistence certificates for themselves and (once
they believe in the encoding) be convinced in the nonexistence of a projective plane
of order ten without having to trust a search procedure.
Second,
using well-tested SAT solvers to perform the
search is less error-prone than writing special-purpose search code%
---a reality of developing software for computer-assisted proofs
is that it is extremely difficult to make custom-written code
both correct and efficient~\cite{lam1990reliable}.

Indeed, our results uncover discrepancies with both the original
1989 search and its 2011 independent verification.  As we detail in
section~\ref{sec:background}, the first two steps of Lam~et~al.'s~search are to enumerate
what
are known as $A1$s and~$A2$s.  Our work agrees with
the previous searches that up to isomorphism there are~66 possibilities
for the $A1$s.  However, our count for the $A2$ possibilities disagrees with
both of the counts of~\cite{lam1989non} and~\cite{roy2011confirmation}%
---works which differ between themselves as well (see Section~\ref{sec:results}).
We generate certificates that demonstrate our $A2$ search is complete
(see Section~\ref{sec:certificates}) and verify the certificates
with a proof verifier.  These certificates were generated
with the assistance of the symbolic computation library
Traces~\cite{McKay201494}---but we describe how one can verify
the certificates without needing to trust the output of the library.

Our work does not provide a completely \emph{formal proof} of the nonexistence of a projective
plane of order ten because we rely on results that currently have no formal
computer-verifiable proofs.
In particular, we rely on a result of~\cite{carter1974existence}
that the error-correcting code associated with a hypothetical projective
plane of order ten must contain words that are referred to as
weight~15, weight~16, or `primitive' weight~19 words.
The former two cases were first ruled out by the searches
of~\cite{macwilliams1973existence} and~\cite{lam1986nonexistence}
and were recently settled via SAT-based nonexistence certificates~\cite{bright2020nonexistence,bright2020unsatisfiability}.
The primitive weight 19
search is by far the most challenging---it was ruled out by~\cite{lam1989non}
and it is the case that we consider in our work.

Our work does provide a possible
avenue for constructing a formal proof: by deriving our SAT encoding
and the mathematical results that we rely on inside a formal proof system.
This would be a significant undertaking but is in principle possible with
current tools.  In fact, results that
were recently proven using SAT solvers such as the resolution of the Boolean Pythagorean
triples problem~\cite{heule2016solving} and a case of the Erd\H{o}s
discrepancy conjecture~\cite{konev2014sat} have since had formal proofs generated based on the
SAT encoding~\cite{cruz2018formally,keller2019}.

\section{Background}\label{sec:background}

We now provide the necessary background in order to understand our results.
In particular, we outline the cube-and-conquer satisfiability solving paradigm,
describe Lam's problem from projective geometry, describe the subcase
for which we provide nonexistence certificates, and outline the symmetry
breaking methods exploited by the nonexistence certificates.

\subsection{Cube-and-conquer}

The \emph{cube-and-conquer} satisfiability solving paradigm was
developed by~\cite{heule2011cube} for solving hard combinatorial
problems.  The method uses two kinds of SAT solvers in two stages:
First, a ``cubing solver'' splits a satisfiability instance into
a large number distinct subproblems specified by \emph{cubes}---formulas
of the form $l_1\land\dotsb\land l_n$ where~$l_i$ are variables
or negated variables.  Second, for each cube a ``conquering solver'' solves the
original instance under the assumption that the cube is true.
The cube-and-conquer method tends to be effective at quickly solving large
satisfiability instances when the cubing solver can generate many cubes
encoding subproblems of approximately equal difficulty.  It has since been applied to
solve huge combinatorial problems such as the Boolean Pythagorean
triples problem~\cite{heule2017solving} and the computation of the fifth
Schur number~\cite{heule2018schur}.

\subsection{Lam's problem}

Projective geometry was formalized by mathematicians in the 1600s though its roots
date back to the work of Pappus of Alexandria in the 4th century.  In the 1800s
projective geometry became extensively studied---including projective geometries
that contain only a finite number of points~\cite{von1856beitrage}.
The only finite projective geometries that remain to be classified are those
in two dimensions and such objects are known as \emph{projective planes}.

Projective planes consist of an incidence relationship
between points and lines such that any two distinct lines intersect in a unique
point and any two distinct points are on a unique line.  To avoid
trivial cases we also require that not all (or all but one) of the points lie
on the same line.  These axioms imply that all lines contain the same number
of points and the plane contains the same number of points and lines.
If each line has $n+1$ points then the projective plane said to be of \emph{order}~$n$
and it will contain exactly $n^2+n+1$ points and the same number of lines.
If $A$ is the $\{0,1\}$ incidence matrix with a $1$ in entry $(i,j)$ exactly
when line~$i$ contains point~$j$
then the projective plane axioms imply that the off-diagonal entries of
the matrices $AA^T$ and $A^TA$ are exactly $1$.  Two vectors are said to
\emph{intersect} if they share a $1$ in the same position and therefore
the projective plane axioms imply that any two rows or columns of $A$ must intersect.

Projective planes are known to exist in all orders that are prime powers
but despite extensive study no projective planes in any other orders
are known and it has been conjectured that the order of a projective plane
must be a prime power~\cite{weisstein2002crc}.  Certain orders including six have
been theoretically ruled out~\cite{bruck1949nonexistence} leaving $n=10$ as the first uncertain
order---until the computational search of~\cite{lam1989non} did not find a plane
of order ten.

Lam et al.'s search was based on properties
of the incidence matrix $A$ of a hypothetical projective plane of order ten.
In particular, the results of~\cite{carter1974existence} and~\cite{macwilliams1973existence}
imply that the words of Hamming weight 19 in the rowspace of $A$ (mod~2)
are of three possible forms (called oval, 16-type, and primitive) and there must exist some
16-type or primitive words.  However, the searches
of~\cite{lam1986nonexistence}
ruled out the existence of 16-type words and~\cite{lam1989non} ruled out
the existence of primitive words.

\subsection{Primitive weight 19 words}

\begin{figure}
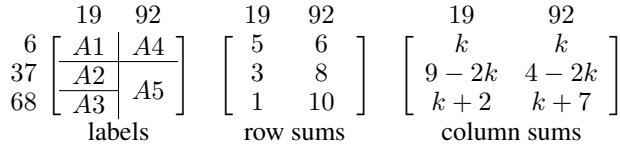

\[
\begin{matrix}
& & 19 & 92 \\
\phantom{0}6\!\!\!\! & \multirow{3}{*}{\rlap{$\left[\rule{0pt}{16pt}\right.$}} & \multicolumn{1}{c|}{\raisebox{-1pt}{$A1$}} & \raisebox{-1pt}{$A4$} & \multirow{3}{*}{\llap{$\left.\rule{0pt}{16pt}\right]$}} \\ \cline{3-4}
37\!\!\!\! & & \multicolumn{1}{c|}{\raisebox{-1pt}{$A2$}} & \multirow{2}{*}{\raisebox{-1pt}{$A5$}} \\ \cline{3-3}
68\!\!\!\! & & \multicolumn{1}{c|}{\raisebox{-1pt}{$A3$}} & \\
& & \multicolumn{2}{c}{\text{labels}\rule{0pt}{8pt}}
\end{matrix}
\quad
\begin{matrix}
& 19 & 92 \\
\multirow{3}{*}{\rlap{$\left[\rule{0pt}{16pt}\right.$}} & 5 & 6 & \multirow{3}{*}{\llap{$\left.\rule{0pt}{16pt}\right]$}} \\
& 3 & 8 \\
& 1 & 10 \\
& \multicolumn{2}{c}{\text{row sums}\rule{0pt}{8pt}}
\end{matrix}
\quad
\begin{matrix}
& 19 & 92 \\
\multirow{3}{*}{\rlap{$\left[\rule{0pt}{16pt}\right.$}} & k & k & \multirow{3}{*}{\llap{$\left.\rule{0pt}{16pt}\right]$}} \\
& 9-2k & 4-2k \\
& k+2 & k+7 \\
& \multicolumn{2}{c}{\text{column sums}\rule{0pt}{8pt}}
\end{matrix}
\]
\caption{Structure of the incidence matrix of a projective plane of order ten
containing a primitive weight 19 word.  The numbers outside the matrices count
the number of rows or columns in each submatrix.
}\label{fig:structure}
\end{figure}

The most challenging case in the resolution of Lam's problem is to
show the nonexistence of primitive weight 19 words.  The existence
of such words greatly constrains the structure of the $111\times111$ incidence matrix~$A$
of a projective plane of order ten---in particular, $A$ can be decomposed
into a $3\times2$ collection of submatrices as shown in Figure~\ref{fig:structure}.
The row sums of the submatrices were derived in~\cite{carter1974existence}
and the column sums (that depend on a parameter~$k$ counting how many~$1$s appear
in the first six rows of a column) were derived in~\cite{lam1985estimates}.
We follow the labelling scheme that appears in the latter work.

Once $A1$ has been fixed this uniquely determines $A3$ and~$A4$ without loss of
generality~\cite{lam1985estimates}.
There are typically a large number of
possibilities for~$A2$, though this number can be reduced by only considering nonisomorphic $A2$s
under the symmetry group of $A1$ (see Section~\ref{subsec:symmetry}).
Once all the possible $A2$s have been determined the search of Lam et~al.~continued by
attempting to extend each $A2$ into the $A5$ submatrix.  In no case was
a completion of $A5$ found, thereby disproving the existence of the complete matrix $A$.

\subsection{Symmetry groups}\label{subsec:symmetry}

Two incidence matrices are said to be \emph{isomorphic} if one can be transformed into
the other through row or column permutations.  The \emph{symmetry group} of a matrix
is the set of row and column permutations that fix all entries of the matrix.
When the symmetry group of a submatrix of~$A$ (such as $A1$ in Figure~\ref{fig:structure}) is large
the search starting from that submatrix tends to include a lot of isomorphic matrices.
To avoid searching through a space containing many isomorphic matrices
an important optimization is to detect and remove as many symmetries as early in the search as possible.

\cite{lam1985estimates} found that up to isomorphism there are exactly 66 possible ways of completing
the submatrix $A1$; an explicit list containing each possible case is
available in~\cite{kaski2006classification}.
Lam et~al.~show how to remove 21 of those possibilities by various argumentation.
Computational searches performed for each of the remaining 45
possibilities~\cite{lam1989non} found 639,624 nonisomorphic ways of extending these $A1$s to $A2$s.
Our count differs from that of
Lam~et~al.---see Section~\ref{sec:results} for details.

\section{SAT encoding}\label{sec:encoding}

We now describe the SAT encoding used in our instances
in three parts---corresponding to the three steps of Lam et~al.'s~search.
First, a single SAT instance determines the possibilities for the $A1$s
(i.e., the upper left $6\times19$ matrix in Figure~\ref{fig:structure}).
Second, for each possible $A1$
a new instance determines the possibilities for the $A2$s.
Third, for each possible $A2$ a new instance determines that
the matrix cannot be completed to a full incidence matrix~$A$.
For efficiency reasons many constraints are dropped from the third set of
instances.  This results in a small number of solutions to these
instances which are then shown to not complete by adding back in
some of the constraints (see Section~\ref{subsec:maincert}).

In each of these instances we use the Boolean variable $a_{i,j}$
to represent that entry $(i,j)$ of $A$ contains a $1$.
Note that since~$A$ defines a projective plane none of its rows
intersect twice; in other words, for every pair of distinct indices
$(i,i')$ there do not exist another pair of distinct indices $(j,j')$
such that all the variables $\{a_{i,j},a_{i,j'},a_{i',j},a_{i',j'}\}$ are true.
Thus, each of our SAT instances include clauses of the form
$\lnot a_{i,j}\lor\lnot a_{i,j'}\lor\lnot a_{i',j}\lor\lnot a_{i',j'}$
where $i$ and $i'$ are indices of the rows under consideration and $j$ and $j'$
are indices of the columns under consideration.

\subsection{\boldmath$A1$ encoding}\label{subsec:a1}

An $A1$ is a $6\times19$ matrix with $\{0,1\}$ entries containing
exactly five $1$s in each row and no two rows having more than
a single~$1$ in the same position.  Furthermore,
the rows and columns of an $A1$ may be lexicographically ordered
without loss of generality~\cite[cf.~exercise 495]{knuth2015art}.
Moreover, lexicographic constraints can simply be encoded into Boolean logic~\cite[cf.~equation 169]{knuth2015art}.
To enforce the fact that the row sum of each row is~5 we use the
sequential counter cardinality encoding~\cite{sinz2005towards}---%
cf.~exercise~30~\cite{knuth2015art}.

In order to find all possible $A1$s we exhaustively search for solutions
of the above SAT instance.  Whenever a solution $S$ is found the \emph{blocking
clause} $\bigvee_{S\models p}\lnot p$ (where $S\models p$
means the variable $p$ is true in $S$) is added to the SAT instance and the
solver is restarted; this produces 3,366 solutions.
To check the $A1$s for equivalence we construct
the \emph{incidence graph} of the matrix~\cite{kaski2006classification}
and use the library Traces~\cite{McKay201494} to discard the $A1$s
whose incidence graphs are isomorphic to those of previously found $A1$s.

\begin{figure}
\centering
\centering
\begin{tikzpicture}[scale=0.35, every node/.style={scale=1}, baseline=(current bounding box.center),ultra thin]
\draw [fill=black] (0,0) rectangle (1,-1) node[midway,white] {\small1};
\draw [fill=black] (1,0) rectangle (2,-1) node[midway,white] {\small1};
\draw [fill=black] (2,0) rectangle (3,-1) node[midway,white] {\small1};
\draw [fill=black] (3,0) rectangle (4,-1) node[midway,white] {\small1};
\draw [fill=black] (4,0) rectangle (5,-1) node[midway,white] {\small1};
\draw (5,0) rectangle (6,-1) node[midway] {\small0};
\draw (6,0) rectangle (7,-1) node[midway] {\small0};
\draw (7,0) rectangle (8,-1) node[midway] {\small0};
\draw (8,0) rectangle (9,-1) node[midway] {\small0};
\draw (9,0) rectangle (10,-1) node[midway] {\small0};
\draw (10,0) rectangle (11,-1) node[midway] {\small0};
\draw (11,0) rectangle (12,-1) node[midway] {\small0};
\draw (12,0) rectangle (13,-1) node[midway] {\small0};
\draw (13,0) rectangle (14,-1) node[midway] {\small0};
\draw (14,0) rectangle (15,-1) node[midway] {\small0};
\draw (15,0) rectangle (16,-1) node[midway] {\small0};
\draw (16,0) rectangle (17,-1) node[midway] {\small0};
\draw (17,0) rectangle (18,-1) node[midway] {\small0};
\draw (18,0) rectangle (19,-1) node[midway] {\small0};
\draw [fill=black] (0,-1) rectangle (1,-2) node[midway,white] {\small1};
\draw (1,-1) rectangle (2,-2) node[midway] {\small0};
\draw (2,-1) rectangle (3,-2) node[midway] {\small0};
\draw (3,-1) rectangle (4,-2) node[midway] {\small0};
\draw (4,-1) rectangle (5,-2) node[midway] {\small0};
\draw [fill=black] (5,-1) rectangle (6,-2) node[midway,white] {\small1};
\draw [fill=black] (6,-1) rectangle (7,-2) node[midway,white] {\small1};
\draw [fill=black] (7,-1) rectangle (8,-2) node[midway,white] {\small1};
\draw [fill=black] (8,-1) rectangle (9,-2) node[midway,white] {\small1};
\draw (9,-1) rectangle (10,-2) node[midway] {\small0};
\draw (10,-1) rectangle (11,-2) node[midway] {\small0};
\draw (11,-1) rectangle (12,-2) node[midway] {\small0};
\draw (12,-1) rectangle (13,-2) node[midway] {\small0};
\draw (13,-1) rectangle (14,-2) node[midway] {\small0};
\draw (14,-1) rectangle (15,-2) node[midway] {\small0};
\draw (15,-1) rectangle (16,-2) node[midway] {\small0};
\draw (16,-1) rectangle (17,-2) node[midway] {\small0};
\draw (17,-1) rectangle (18,-2) node[midway] {\small0};
\draw (18,-1) rectangle (19,-2) node[midway] {\small0};
\draw [fill=black] (0,-2) rectangle (1,-3) node[midway,white] {\small1};
\draw (1,-2) rectangle (2,-3) node[midway] {\small0};
\draw (2,-2) rectangle (3,-3) node[midway] {\small0};
\draw (3,-2) rectangle (4,-3) node[midway] {\small0};
\draw (4,-2) rectangle (5,-3) node[midway] {\small0};
\draw (5,-2) rectangle (6,-3) node[midway] {\small0};
\draw (6,-2) rectangle (7,-3) node[midway] {\small0};
\draw (7,-2) rectangle (8,-3) node[midway] {\small0};
\draw (8,-2) rectangle (9,-3) node[midway] {\small0};
\draw [fill=black] (9,-2) rectangle (10,-3) node[midway,white] {\small1};
\draw [fill=black] (10,-2) rectangle (11,-3) node[midway,white] {\small1};
\draw [fill=black] (11,-2) rectangle (12,-3) node[midway,white] {\small1};
\draw [fill=black] (12,-2) rectangle (13,-3) node[midway,white] {\small1};
\draw (13,-2) rectangle (14,-3) node[midway] {\small0};
\draw (14,-2) rectangle (15,-3) node[midway] {\small0};
\draw (15,-2) rectangle (16,-3) node[midway] {\small0};
\draw (16,-2) rectangle (17,-3) node[midway] {\small0};
\draw (17,-2) rectangle (18,-3) node[midway] {\small0};
\draw (18,-2) rectangle (19,-3) node[midway] {\small0};
\draw [fill=black] (0,-3) rectangle (1,-4) node[midway,white] {\small1};
\draw (1,-3) rectangle (2,-4) node[midway] {\small0};
\draw (2,-3) rectangle (3,-4) node[midway] {\small0};
\draw (3,-3) rectangle (4,-4) node[midway] {\small0};
\draw (4,-3) rectangle (5,-4) node[midway] {\small0};
\draw (5,-3) rectangle (6,-4) node[midway] {\small0};
\draw (6,-3) rectangle (7,-4) node[midway] {\small0};
\draw (7,-3) rectangle (8,-4) node[midway] {\small0};
\draw (8,-3) rectangle (9,-4) node[midway] {\small0};
\draw (9,-3) rectangle (10,-4) node[midway] {\small0};
\draw (10,-3) rectangle (11,-4) node[midway] {\small0};
\draw (11,-3) rectangle (12,-4) node[midway] {\small0};
\draw (12,-3) rectangle (13,-4) node[midway] {\small0};
\draw [fill=black] (13,-3) rectangle (14,-4) node[midway,white] {\small1};
\draw [fill=black] (14,-3) rectangle (15,-4) node[midway,white] {\small1};
\draw [fill=black] (15,-3) rectangle (16,-4) node[midway,white] {\small1};
\draw [fill=black] (16,-3) rectangle (17,-4) node[midway,white] {\small1};
\draw (17,-3) rectangle (18,-4) node[midway] {\small0};
\draw (18,-3) rectangle (19,-4) node[midway] {\small0};
\draw (0,-4) rectangle (1,-5) node[midway] {\small0};
\draw [fill=black] (1,-4) rectangle (2,-5) node[midway,white] {\small1};
\draw (2,-4) rectangle (3,-5) node[midway] {\small0};
\draw (3,-4) rectangle (4,-5) node[midway] {\small0};
\draw (4,-4) rectangle (5,-5) node[midway] {\small0};
\draw [fill=black] (5,-4) rectangle (6,-5) node[midway,white] {\small1};
\draw (6,-4) rectangle (7,-5) node[midway] {\small0};
\draw (7,-4) rectangle (8,-5) node[midway] {\small0};
\draw (8,-4) rectangle (9,-5) node[midway] {\small0};
\draw [fill=black] (9,-4) rectangle (10,-5) node[midway,white] {\small1};
\draw (10,-4) rectangle (11,-5) node[midway] {\small0};
\draw (11,-4) rectangle (12,-5) node[midway] {\small0};
\draw (12,-4) rectangle (13,-5) node[midway] {\small0};
\draw [fill=black] (13,-4) rectangle (14,-5) node[midway,white] {\small1};
\draw (14,-4) rectangle (15,-5) node[midway] {\small0};
\draw (15,-4) rectangle (16,-5) node[midway] {\small0};
\draw (16,-4) rectangle (17,-5) node[midway] {\small0};
\draw [fill=black] (17,-4) rectangle (18,-5) node[midway,white] {\small1};
\draw (18,-4) rectangle (19,-5) node[midway] {\small0};
\draw (0,-5) rectangle (1,-6) node[midway] {\small0};
\draw [fill=black] (1,-5) rectangle (2,-6) node[midway,white] {\small1};
\draw (2,-5) rectangle (3,-6) node[midway] {\small0};
\draw (3,-5) rectangle (4,-6) node[midway] {\small0};
\draw (4,-5) rectangle (5,-6) node[midway] {\small0};
\draw (5,-5) rectangle (6,-6) node[midway] {\small0};
\draw [fill=black] (6,-5) rectangle (7,-6) node[midway,white] {\small1};
\draw (7,-5) rectangle (8,-6) node[midway] {\small0};
\draw (8,-5) rectangle (9,-6) node[midway] {\small0};
\draw (9,-5) rectangle (10,-6) node[midway] {\small0};
\draw [fill=black] (10,-5) rectangle (11,-6) node[midway,white] {\small1};
\draw (11,-5) rectangle (12,-6) node[midway] {\small0};
\draw (12,-5) rectangle (13,-6) node[midway] {\small0};
\draw (13,-5) rectangle (14,-6) node[midway] {\small0};
\draw [fill=black] (14,-5) rectangle (15,-6) node[midway,white] {\small1};
\draw (15,-5) rectangle (16,-6) node[midway] {\small0};
\draw (16,-5) rectangle (17,-6) node[midway] {\small0};
\draw (17,-5) rectangle (18,-6) node[midway] {\small0};
\draw [fill=black] (18,-5) rectangle (19,-6) node[midway,white] {\small1};
\end{tikzpicture}
\centering
\begin{tikzpicture}[scale=0.12, every node/.style={scale=1}, baseline=(current bounding box.center),ultra thin]
\node at (-1.5,-0.5) {\tiny7};
\node at (-1.5,-1.5) {\tiny8};
\node at (-1.5,-4.5) {\tiny11};
\node at (-1.5,-11.5) {\tiny18};
\node at (-1.5,-18.5) {\tiny25};
\draw [fill=black] (0,0) rectangle (1,-1);
\draw (1,0) rectangle (2,-1);
\draw (2,0) rectangle (3,-1);
\draw (3,0) rectangle (4,-1);
\draw (4,0) rectangle (5,-1);
\draw (0,-1) rectangle (1,-2);
\draw [fill=black] (1,-1) rectangle (2,-2);
\draw (2,-1) rectangle (3,-2);
\draw (3,-1) rectangle (4,-2);
\draw (4,-1) rectangle (5,-2);
\draw (0,-2) rectangle (1,-3);
\draw [fill=black] (1,-2) rectangle (2,-3);
\draw (2,-2) rectangle (3,-3);
\draw (3,-2) rectangle (4,-3);
\draw (4,-2) rectangle (5,-3);
\draw (0,-3) rectangle (1,-4);
\draw [fill=black] (1,-3) rectangle (2,-4);
\draw (2,-3) rectangle (3,-4);
\draw (3,-3) rectangle (4,-4);
\draw (4,-3) rectangle (5,-4);
\draw (0,-4) rectangle (1,-5);
\draw (1,-4) rectangle (2,-5);
\draw [fill=black] (2,-4) rectangle (3,-5);
\draw (3,-4) rectangle (4,-5);
\draw (4,-4) rectangle (5,-5);
\draw (0,-5) rectangle (1,-6);
\draw (1,-5) rectangle (2,-6);
\draw [fill=black] (2,-5) rectangle (3,-6);
\draw (3,-5) rectangle (4,-6);
\draw (4,-5) rectangle (5,-6);
\draw (0,-6) rectangle (1,-7);
\draw (1,-6) rectangle (2,-7);
\draw [fill=black] (2,-6) rectangle (3,-7);
\draw (3,-6) rectangle (4,-7);
\draw (4,-6) rectangle (5,-7);
\draw (0,-7) rectangle (1,-8);
\draw (1,-7) rectangle (2,-8);
\draw [fill=black] (2,-7) rectangle (3,-8);
\draw (3,-7) rectangle (4,-8);
\draw (4,-7) rectangle (5,-8);
\draw (0,-8) rectangle (1,-9);
\draw (1,-8) rectangle (2,-9);
\draw [fill=black] (2,-8) rectangle (3,-9);
\draw (3,-8) rectangle (4,-9);
\draw (4,-8) rectangle (5,-9);
\draw (0,-9) rectangle (1,-10);
\draw (1,-9) rectangle (2,-10);
\draw [fill=black] (2,-9) rectangle (3,-10);
\draw (3,-9) rectangle (4,-10);
\draw (4,-9) rectangle (5,-10);
\draw (0,-10) rectangle (1,-11);
\draw (1,-10) rectangle (2,-11);
\draw [fill=black] (2,-10) rectangle (3,-11);
\draw (3,-10) rectangle (4,-11);
\draw (4,-10) rectangle (5,-11);
\draw (0,-11) rectangle (1,-12);
\draw (1,-11) rectangle (2,-12);
\draw (2,-11) rectangle (3,-12);
\draw [fill=black] (3,-11) rectangle (4,-12);
\draw (4,-11) rectangle (5,-12);
\draw (0,-12) rectangle (1,-13);
\draw (1,-12) rectangle (2,-13);
\draw (2,-12) rectangle (3,-13);
\draw [fill=black] (3,-12) rectangle (4,-13);
\draw (4,-12) rectangle (5,-13);
\draw (0,-13) rectangle (1,-14);
\draw (1,-13) rectangle (2,-14);
\draw (2,-13) rectangle (3,-14);
\draw [fill=black] (3,-13) rectangle (4,-14);
\draw (4,-13) rectangle (5,-14);
\draw (0,-14) rectangle (1,-15);
\draw (1,-14) rectangle (2,-15);
\draw (2,-14) rectangle (3,-15);
\draw [fill=black] (3,-14) rectangle (4,-15);
\draw (4,-14) rectangle (5,-15);
\draw (0,-15) rectangle (1,-16);
\draw (1,-15) rectangle (2,-16);
\draw (2,-15) rectangle (3,-16);
\draw [fill=black] (3,-15) rectangle (4,-16);
\draw (4,-15) rectangle (5,-16);
\draw (0,-16) rectangle (1,-17);
\draw (1,-16) rectangle (2,-17);
\draw (2,-16) rectangle (3,-17);
\draw [fill=black] (3,-16) rectangle (4,-17);
\draw (4,-16) rectangle (5,-17);
\draw (0,-17) rectangle (1,-18);
\draw (1,-17) rectangle (2,-18);
\draw (2,-17) rectangle (3,-18);
\draw [fill=black] (3,-17) rectangle (4,-18);
\draw (4,-17) rectangle (5,-18);
\draw (0,-18) rectangle (1,-19);
\draw (1,-18) rectangle (2,-19);
\draw (2,-18) rectangle (3,-19);
\draw (3,-18) rectangle (4,-19);
\draw [fill=black] (4,-18) rectangle (5,-19);
\draw (0,-19) rectangle (1,-20);
\draw (1,-19) rectangle (2,-20);
\draw (2,-19) rectangle (3,-20);
\draw (3,-19) rectangle (4,-20);
\draw [fill=black] (4,-19) rectangle (5,-20);
\draw (0,-20) rectangle (1,-21);
\draw (1,-20) rectangle (2,-21);
\draw (2,-20) rectangle (3,-21);
\draw (3,-20) rectangle (4,-21);
\draw [fill=black] (4,-20) rectangle (5,-21);
\draw (0,-21) rectangle (1,-22);
\draw (1,-21) rectangle (2,-22);
\draw (2,-21) rectangle (3,-22);
\draw (3,-21) rectangle (4,-22);
\draw [fill=black] (4,-21) rectangle (5,-22);
\draw (0,-22) rectangle (1,-23);
\draw (1,-22) rectangle (2,-23);
\draw (2,-22) rectangle (3,-23);
\draw (3,-22) rectangle (4,-23);
\draw [fill=black] (4,-22) rectangle (5,-23);
\draw (0,-23) rectangle (1,-24);
\draw (1,-23) rectangle (2,-24);
\draw (2,-23) rectangle (3,-24);
\draw (3,-23) rectangle (4,-24);
\draw [fill=black] (4,-23) rectangle (5,-24);
\draw (0,-24) rectangle (1,-25);
\draw (1,-24) rectangle (2,-25);
\draw (2,-24) rectangle (3,-25);
\draw (3,-24) rectangle (4,-25);
\draw [fill=black] (4,-24) rectangle (5,-25);
\end{tikzpicture}
\caption{On the left the $A1$ of the first case is shown.
On the right the upper-left
$25\times5$ submatrix of its associated~$A2$ is shown with some of its row numbers in $A$---this submatrix is
completely determined under the assumption that
its rows are lexicographically ordered.}\label{fig:a1a2}
\end{figure}
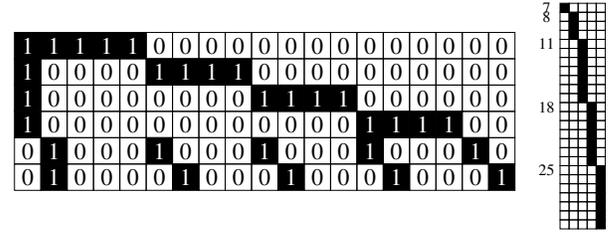

\subsection{\boldmath$A2$ encoding}\label{subsec:a2}

An $A2$ is a $37\times19$ matrix with $\{0,1\}$ entries containing exactly
three $1$s in each row and no two rows having more than
a single~$1$ in the same location.  Furthermore, in the complete incidence matrix $A$ the $A2$ appears
directly below an $A1$ and this completely determines the number of $1$s
that appear in each column (see Figure~\ref{fig:structure}).
Because the rows of $A2$ may be permuted freely without loss of generality
we assume that the rows of $A2$ are lexicographically sorted.
The row sum, column sum, and lexicographic constraints are each encoded
in the same way as in the $A1$ encoding.

The $A2$ SAT instances may now be exhaustively solved
similar to how the $A1$ SAT instance is solved and this
produces over~56~million solutions.
Due to the large number of solutions
it is much more efficient to employ symmetry removal
during the solving process.  To this end we adapt the ``recorded
objects'' method of isomorph-free exhaustive
generation~\cite{kaski2006classification} to the SAT context.
In order to do this we first split the $A2$ search space into a number
of successive levels, where each level successively fills in
more of the $A2$ matrix.

Note that the first row of $A1$ is $1^50^{14}$ (as a binary vector)
because its columns are lexicographically ordered.
Thus the first five columns pairwise intersect in $A1$ and
so will not intersect at all in $A2$.  It follows that the column sum
and lexicographic constraints completely fix the first five columns
of~$A2$.  For example, Figure~\ref{fig:a1a2} contains one possible
$A1$ and the first five columns of the $A2$ generated by this $A1$.
The set of rows of $A2$ can be split into \emph{levels} based on
which rows are identical in the first five columns.  For example, the levels
of the $A2$ in Figure~\ref{fig:a1a2} consist of row 7 and then
the rows 8--10, 11--17, 18--24, 25--31, and the remaining rows 32--43
(consisting of zeros in the first five columns).

Our exhaustive search proceeds by finding completions of the entries
in each successive level.  At each level isomorphism removal is performed on the
completions found at that level.
When multiple completions are isomorphic to each
other only a single one of those completions is used
for the purposes of completing the next level.

When a completion of a level is found, the
incidence graph of the $A1$
and $A2$ (up to the given level) is formed and passed to the
Traces library.  Traces
generates a certificate of the incidence graph and if the certificate
is new the completion is \emph{recorded} as a new solution of the current
level.  If the certificate has been previously seen, the incidence
graph is compared to the incidence graphs of the previously
found completions to verify that it is indeed isomorphic to one of them.
Once it has been verified that the completion is isomorphic to a previously
found completion the new completion is discarded.
Formally, a blocking clause $\bigvee_{S\models p}\lnot p$ is added
to the SAT instance, where $S$ consists of the assignment
formed by the completion to be discarded (up to the given level).
The solver is then resumed---%
the blocking clause being added to the SAT instance on-the-fly, i.e.,
\emph{programmatically}~\cite{ganesh2012lynx}.

This procedure requires keeping track of the nonisomorphic completions
(and their certificates) that
are found at each level as the search is progressing but its advantage
is that symmetries are detected and removed much earlier than they
would otherwise be if isomorphism removal was only performed at the final level.
Moreover, the search is still exhaustive in the sense that all
nonisomorphic $A2$s will be found.
Formally, we state the following theorem whose proof appears in the
online appendix.

\begin{theorem}\label{thm:record}
If the $A2$ SAT instances are solved with isomorphism removal performed
after the completion of each level then
the solver will record exactly one
representative from each equivalence class
of $A2$ completions.
\end{theorem}

\subsection{Main encoding}\label{subsec:main}

\begin{figure}
\input{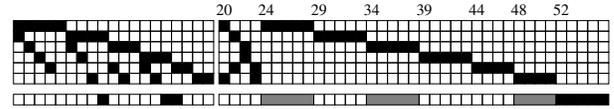}
\caption{The $A1$ of case 66 (upper left) and its associated $A4$ (upper right).
One particular row of an $A2$ is shown (lower left) and the form of
its completion in $A5$ is shown (lower right).  Each of the gray
rectangles contain a single~$1$ and there are another five $1$s
on the row---ordered here to appear as soon as possible.}\label{fig:a1a4}
\end{figure}

We now describe our main encoding of determining the nonexistence
of $A$ containing a given $A1$ and $A2$.  First, 
there is a unique way of completing $A3$ (assuming a lexicographic ordering of its rows)
because its row sums are~$1$.  Similarly,
there is a unique way of completing $A4$ because its columns
have at most two $1$s (having $k\geq3$ in the rightmost column
of Figure~\ref{fig:structure} is impossible).
We complete the columns of $A4$ with column sum~$2$ first (adding
a single column of this form for each pair of lines that do not already
intersect in the first 19 columns) and assume a lexicographic
ordering of the remaining columns.
Figure~\ref{fig:a1a4} shows one possible $A1$ and its associated $A4$.

There are six \emph{blocks} appearing in an $A4$
where block~$i$ consists of the columns containing a $1$ in
the~$i$th row.
The columns not incident to any of the blocks
are called \emph{outside} columns.
For example, in Figure~\ref{fig:a1a4} the first block
consists of column 20 with columns 24--28, and the outside columns
are columns 52--56.

The columns of $A5$ that are identical in $A4$ can be taken
to appear in lexicographic order without loss of generality---in
other words, we assume that the columns within each block
(except for those with column sum~$2$ in~$A4$)
are lexicographically sorted.
Similarly, the rows in~$A5$ that are identical in~$A3$ are
assumed to appear in lexicographic order.

At this point we could simply encode the row sums, column sums,
and lexicographic constraints using the same encoding that
was used in the $A1$ and $A2$ encodings.
The work of Lam et~al.~imply these instances
are unsatisfiable, but we optimized the
encoding in order to reduce the amount of computational resources required
to prove unsatisfiability.

In particular, we did not use all 92 columns of $A5$ because this resulted in an
excessive number of constraints.  Instead, we selected
between 4 and 6 blocks (see Section~\ref{subsec:block})
from each instance and only used the constraints appearing
in those blocks.  In this context
the row sum constraints no longer apply and we ignore
the column sum constraints in the bottom (last 68 rows) of $A5$.
Instead, we use an alternative encoding that directly
enforces incidences that the first~19 columns have in $A5$.
We also use an improved lexicographic encoding taking into account
the form of the rows or columns being ordered.

\subsubsection{Incidence constraints}

The first 19 columns of $A$ are known in each SAT instance;
say $C_j$ denotes the set of row indices
of the $1$s in column~$j$ with $1\leq j\leq19$.
The axioms of a projective plane imply that any two
columns intersect---thus for each pair of column indices $(j,k)$
with $1\leq j\leq19$ and $k$ in the
blocks we are completing we include the clause $\bigvee_{i\in C_j} a_{i,k}$.
Similarly, if $R_i$ denotes the set of column indices
of the $1$s in row~$i$ we include clauses of the form $\bigvee_{j\in R_i} a_{k,j}$
where $7\leq k\leq111$ and block~$i$ is selected.

\subsubsection{Lexicographic constraints}

Consider the columns of a block that have a single~$1$ in $A4$
(e.g., columns 24--28 in Figure~\ref{fig:a1a4}).
By Figure~\ref{fig:structure} we know there are exactly
three $1$s in the first 43 rows of these columns (two
of which appear in $A5$).  If $i$ and $i'$ are the row
indices of the $1$s in column $j$ that is not
the final column in a block then the lexicographic
ordering of the columns implies that $A_{i^*,j+1}=0$
for all $7\leq i^*<\min(i,i')$.  Translating this into
conjunctive normal form, we include the clauses
$\lnot a_{i,j}\lor\lnot a_{i',j}\lor\lnot a_{i^*,j+1}$
for all $7\leq i^*<i'<i\leq 43$ and all columns $j$ that
have a single~$1$ in $A4$ (except for the final column in a block).
This generalizes the lexicographic encoding of~\cite{bright2020unsatisfiability};
and we encoded the row lexicographic constraints in a similar way.

These constraints uniquely fix certain entries.  For example,
if the bottom row of Figure~\ref{fig:a1a4} was the first row
in $A2$ then the leftmost entry in each gray rectangle would
have to be~$1$.  In our main instances we reorder the rows
of $A2$ to maximize the number of mutually intersecting rows
at the top in order to fix as many entries as possible.

\subsubsection{Partial isomorphism removal constraints}

These constraints are optional and tend to only apply in the
easiest cases, though when they do apply
they effectively constrain the search.
They encode the ``extra partial
isomorphism testing'' condition of~\cite{lam1989non}.
Briefly, if two of the first six rows intersect after the first 19
columns then the vector formed by adding (mod 2) these rows
to $1^{19}0^{92}$ (as a binary vector) forms a new vector
that intersects six rows of $A$ five times each.  The specific six rows
vary between cases but may quickly
be determined in each case by examining $A1$ and $A2$.  There
are only a small number of ways of completing those six rows and each
way must be isomorphic to one of the 66 possibilities for $A1$.
If the completion corresponds to a case that has already been solved
then a clause blocking that completion can be
included in the SAT instance as it cannot lead to a solution.
If the $A1$ cases are ordered so that those with the fewest number
of $A2$s are solved first then about~%
22\% of~$A2$s have all of their completions isomorphic
to previously solved cases.

\subsection{Block selection}\label{subsec:block}

Each of our SAT instances used the columns
from~4 to~6 blocks.  We used two different methods of selecting the
blocks: a simple method which was used in the easiest cases and a
more involved method which was used in the hardest cases and
which we experimentally found was about 4 to 5 times faster
in those cases.

\subsubsection{Inside block method}

This method was used on the cases with two or more columns with
column sum $2$ in $A4$---these tend to be the easiest cases.
This method simply selects five blocks (i.e., ignores a single block)
chosen to maximize the number of intersections that occur
in the first six rows of the selected blocks.
For example, in Figure~\ref{fig:a1a4} blocks 5 and 6 have the most
intersections in $A4$ so they are chosen.  Additionally,
blocks 1 and 2 are chosen since they intersect blocks 5 and 6.
If possible, the block that is ignored is a block that
has no intersections in $A4$; otherwise, one of the remaining
blocks with a single intersection in $A4$ is ignored
(e.g., block 3 in Figure~\ref{fig:a1a4}).

\subsubsection{Outside block method}

This method was used on the cases with at most a single
column with column sum $2$ in $A4$---these tend to be the hardest cases
because of the lack of intersections between blocks in $A4$ (which
tend to produce conflicts leading to quick proofs of unsatisfiability).
Thus, in these cases we used a seventh ``outside''
block that shares columns with as many of the first six (or \emph{inside}) blocks as possible.

First, we choose a ``special'' line in the first 37 rows of $A5$ which includes points
in as many inside blocks as possible.  For example, in Figure~\ref{fig:a1a4}
the line at the bottom includes points from the blocks 1, 3, and~6,
though in some cases a line can be found that is incident to all six inside blocks.
The remaining points on the special line become the outside block.
For example, in Figure~\ref{fig:a1a4} the outside block
would include the last five columns of the diagram
if the bottom line was selected as the special line.

The SAT instance only uses blocks incident to the special line
and the blocks (if any) that intersect in the first six rows.
Additionally, in order to reduce the number of constraints one inside
block is dropped from the instance (so long as
at least four blocks in total are selected).  The dropped block
is selected to have the fewest number of rows
with unassigned entries in common with the other blocks.  Such a 
block is the least likely to produce conflicts---since conflicts
between two blocks occur in rows where both blocks have unassigned
entries.

\section{Certificates}\label{sec:certificates}

In this section we describe the certificates from our resolution of Lam's problem
and how a third party can verify them.  We use several kinds of certificates
based on the different parts of the search.  First, we provide a certificate which specifies
the possible cases for $A1$.  For each $A1$ we provide another
certificate specifying each of the ways (if any) of extending that $A1$ to $A2$,
and similarly for each $A2$ we provide a certificate specifying each of the ways (if any)
of extending that $A2$ to the selected blocks
(as described in Section~\ref{subsec:block}) of $A5$.  Finally, we provide certificates
showing all completions of the selected blocks do not extend to a complete $A$
and thereby resolve Lam's problem.
Our certificates are based on the DRAT (deletion
resolution asymmetric tautology) format~\cite{wetzler2014drat}
which is a standard format used for verifying the unsatisfiability of SAT instances.

\subsection{\boldmath$A1$ certificate}\label{subsec:a1certificate}

The $A1$ certificate consists of a DRAT proof that the $A1$ SAT instance adjoined with
3,366 blocking clauses (one for each solution of the original
instance) is unsatisfiable, thus providing all solutions of the
original instance.  Furthermore, a separate certificate (produced by the library Traces)
contains an explicit permutation of the
rows and columns of each solution that produces
a unique canonical form in each case.
Using these permutations one can verify
each solution is isomorphic to one of 66 $A1$s, providing an upper bound on the
number of $A1$s that need to be considered---all that is strictly necessary
to verify the resolution of Lam's problem.

Traces also provides the elements of the symmetry group of each $A1$.
One can verify this output without needing to trust Traces' implementation;
it is straightforward to verify that the permutations
produced by Traces do in fact fix the entries of $A1$ and one can independently
check (e.g., by hand) that the size of the symmetry group is correct.

\subsection{\boldmath$A2$ certificates}\label{subsec:a2certificate}

The $A2$ certificates consist of DRAT proofs of unsatisfiability for
each $A2$ SAT instance adjoined with blocking clauses
for every solution found and blocking clauses for every
(possibly partial) completion that is isomorphic to
a recorded completion (as described in Section~\ref{subsec:a2}).
Accompanying each solution and
partial completion is a set of row and column permutations (provided
by Traces) for
translating the solution or partial completion into a canonical form.

In order to verify one of the certificates, one needs to verify that
(1) the DRAT proof does indeed show the unsatisfiability of the
augmented SAT instance, and (2) the canonical form of each
(partial or complete) solution is
equal to the canonical form of one of the recorded nonisomorphic completions.

By Theorem~\ref{thm:record} this shows that the set of solutions
consists of exactly one solution from each equivalence class of
solutions to the original SAT instance.  Moreover, each blocked
clause is independently justified
(without trusting Traces)
to block one of the recorded solutions or to block
a partial completion isomorphic to a recorded partial completion.

The fact that each of the recorded solutions are nonisomorphic
to each other \emph{does} rely on trusting Traces' canonical form.
In order to verify the resolution
of Lam's problem it is not strictly necessary to verify
the recorded $A2$s are nonisomorphic---%
however, in order to verify (without trusting Traces) that
our $A2$ counts did not include extraneous cases we also performed a verification
that all recorded solutions were nonisomorphic of each other.

Note that all $A2$s that are isomorphic to a given $A2$ can be generated
through the symmetry group of its associated $A1$;
up to row permutations any two isomorphic
$A2$s must be isomorphic to each other via a permutation in the symmetry
group of $A1$.
Additionally, as mentioned in Section~\ref{subsec:a1certificate}, the
symmetry group of each $A1$ can be verified without trusting Traces.
No two recorded $A2$s are isomorphic to each other under
the symmetries of their associated $A1$,
thereby showing each of the recorded $A2$s are indeed mutually nonisomorphic.
This can be exhaustively verified by applying
every $A1$ symmetry group permutation to every recorded $A2$.

\subsection{Main certificates}\label{subsec:maincert}

The main certificates consist of DRAT proofs of unsatisfiability for
each of the SAT instances described in Section~\ref{subsec:main}
adjoined with blocking clauses for the completions (if any)
of the blocks selected to appear in the SAT instance.
The majority of SAT instances had no completions but in about~3\%
of them completions were found---%
see the online appendix for one explicit such completion.

An individual certificate for each completion can be generated
showing that the completion does not extend to a complete incidence
matrix~$A$.  Alternatively, it is more efficient to generate one
certificate for each case $A1$ showing that none of the completions found
in that case can be extended.  In order to do this, we generate
a SAT instance for each $A1$ that includes the variables and constraints from all six inside blocks
and those from the outside block columns appearing in each completion.
Each completion $C$ found for that $A1$
is specified as a set of incremental assumption
unit clauses~\cite{audemard2013improving}.
Once the solver finds no solutions while assuming
$\bigwedge_{C\models p}p$ it moves on to the next set of assumptions.
This ``incremental'' solve produces a single DRAT certificate which
proves the clause $\bigvee_{C\models p}\lnot p$ for each completion~$C$
under consideration---thereby demonstrating that these
completions~$C$ cannot be extended to all six inside blocks.
This verifies that a full completion of $A$ cannot in fact exist.

\section{Results}\label{sec:results}

Our SAT instances are generated by Python scripts that
are freely available as a part of the ``MathCheck'' project.  The
code, along with Bash scripts to generate and check the
certificates, is available from \url{uwaterloo.ca/mathcheck}.
Unless otherwise specified we solve the SAT
instances using MapleSAT~\cite{liang2016exponential}
and verify the certificates using
GRATgen~\cite{lammich2017efficient}.
The computations were performed on a cluster of Intel~E5
cores at 2.1~GHz running Linux and using at most 4GB of memory.

\subsection{\boldmath$A1$ and $A2$ results}

The $A1$ instance can be generated and solved
in a few seconds.  The resulting
certificate (about 1MB) shows that there are 3,366 total
solutions of the SAT instance and 3,300 of them are
isomorphic to one of the remaining 66 $A1$s.
We label these $A1$s using the case numbers given
in~\cite{kaski2006classification}.

The $A2$ instances are generated and solved
in about 25 minutes and produce a total of 650,370 nonisomorphic
$A2$s.
The resulting certificates
along with the canonical form labellings provided by Traces
total about 7GB.  Fifteen cases (4, 10, 14, 19, 28--31, 35, 40, 44, 45, 59,
61, and 62) are found to have no $A2$s.
Five cases (32, 38, 54, 57, and~64) imply the existence
of a word of weight~16 in $A$'s rowspace~\cite{lam1985estimates}
and previous searches~\cite{carter1974existence,lam1986nonexistence}
and certificates~\cite{bright2020unsatisfiability}
demonstrates the nonexistence of such a word.  Case 52 is
eliminated by a theoretical argument~\cite{lam1985estimates},
leaving 45 cases remaining.

Our counts for the number of nonisomorphic $A2$s in the remaining 45 cases
match those of~\cite{lam1989non} in all but the eight cases
shown in Table~\ref{tbl:counts}.  The independent
verification~\cite{roy2011confirmation}
does not provide the number of $A2$s found in each case
but their total $A2$ count is inconsistent
with both the counts of Lam et~al.~and the counts
provided by our certificates.
It is unclear what caused the discrepancy between these
searches---the previous searches did not produce certificates
and by personal communication we have been informed
that the code and data from the
previous searches are no longer available.

\begin{table}
\[\begin{tabular}{ccc@{\qquad}ccc}
Case & Lam & Our Work & Case & Lam & Our Work \\
11 & \phantom{0}7,397 & \phantom{0}7,059 & 39 & 1,010 & \phantom{0,}505 \\
12 & 10,966 & 10,635 & 56 & 1,554 & \phantom{0,}794 \\
16 & \phantom{0}5,958 & \phantom{0}8,040 & 58 & 4,329 & 4,188 \\
23 & \phantom{0}8,033 & \phantom{0}7,971 & 66 & \phantom{0,}662 & \phantom{0,}168 
\end{tabular}\]
\caption{The $A2$ counts given by~\cite{lam1989non}
that differ with the counts given by our work.}\label{tbl:counts}
\end{table}

The method described in Section~\ref{subsec:a2certificate}
of checking that the generated $A2$s are mutually
nonisomorphic generates 56,157,420 total $A2$s;
complete counts for each case is available in the online appendix.  Adding
these $A2$s as blocking clauses to the original $A2$ instances (without using symmetry removal)
produces unsatisfiable instances that
provide a second verification that no $A2$s were missed%
---and without relying on Theorem~\ref{thm:record}.  However,
the instances generated in this way took
about 150 times longer to solve.

\subsection{Main results}

The 45 remaining cases have a total of 639,075 nonisomorphic $A2$s
between them.  SAT instances are generated for each of these
and are simplified using the solver \textsc{CaDiCaL}~\cite{biere2019cadical}
run for 20,000 conflicts.  In total this simplification
uses about 400 hours and determines that 166,408 of the instances are unsatisfiable.

The remaining simplified instances are
processed using the ``cubing'' solver March\_cu~\cite{heule2011cube}.
We disabled the default cubing cutoff of this solver in favour
of a cutoff based on the number of free variables in the subproblems
specified by each cube.  More precisely, when the number of free
variables in a subproblem drops below a provided bound no more
cubing occurs in that subproblem.
The cubing bound is controlled
by March\_cu's \texttt{-n} parameter, but we modified March\_cu so that
the auxiliary variables from the cardinality
constraints are not considered free.  The cutoff bound
was experimentally chosen by randomly selecting up to several hundred
instances from each case and determining a bound that minimizes
the sum of the cubing and conquering times.  Ultimately, the
cubing solver produces over~312~million cubes
and uses about 1,200 hours.

The simplified SAT instances are solved using the
conquering solver MapleSAT with the cubes produced by March\_cu.
This requires about 15,000 total core hours or about 16 hours of
real time when simultaneously distributed across 30 machines with 32 cores each.
In each instance the DRAT proof produced by MapleSAT is concatenated
with the simplification proof produced by \textsc{CaDiCaL}.
The combined proofs total about 110 tebibytes in a binary format and are used to verify
that each of the original SAT instances (after adding blocking
clauses for each solution found by MapleSAT) are unsatisfiable;
GRATgen checked the proofs in about 33,000 core hours.
It is possible that each solution found by MapleSAT leads to more than
one solution of the original SAT instance because
variables are eliminated during the simplification process.
Regardless, the DRAT proofs show that the solutions found by MapleSAT
are exhaustive in the sense that any satisfying assignment of the original SAT instance
must extend a solution found by MapleSAT.

MapleSAT finds 24,882 partial solutions and
these are all shown to not extend to a full incidence matrix
in a total of about 6 minutes.  These certificates are about 4GB and are
checked with DRAT-trim~\cite{wetzler2014drat}
in forward checking mode.  The default backward
checking mode can not be used as these certificates
are generated using incremental assumptions and therefore
prove the negation of each set of assumptions rather than an empty clause.

\section{Conclusion}

In this paper we have completed a resolution of Lam's problem
from finite geometry---the problem of showing the nonexistence
of a projective plane of order ten.  Extensive searches
solved this problem in a landmark result in the 1980s and
this result remains one of the most significant results in computational
combinatorial classification.
Our work improves on these searches by producing certificates
that can be verified by a third party using a proof verifier.

In contrast to the previous resolutions of Lam's problem our
work is less error-prone in the sense that it does not require
writing custom-purpose search algorithms.  Instead, we
reduce the problem to SAT and use SAT solvers to perform all
of the exhaustive searches.  Our work demonstrates
the benefits of this approach, as we uncover inconsistencies
with both the original search and an independent confirmation.

Moreover, our search provides the fastest known demonstration
of the nonexistence of primitive weight 19 words in a projective
plane of order ten (in part due to
increases in computational capacity).  Our search shows
this nonexistence result using about 24 months
on desktop CPUs (at~2.1~GHz), while~\cite{roy2011confirmation} used about
27 months on desktop CPUs (at~2.4~GHz)
and~\cite{lam1989non} used about 27 months
on a VAX 11/780 and about 3 months on a CRAY-1A.

As previously mentioned our work does not provide
a \emph{formal} proof resolving Lam's problem because it relies
on some theoretical results that currently have no computer
verifiable proofs as well as some unverified scripts that generate
and solve the SAT instances.  As future work we would like
to see a completely formal verification based on a SAT encoding.
This will likely be a significant challenge due to the 
amount of theoretical results that the encoding relies on
as well as the extensive parallelization that is seemingly necessary
in order to check the proof in a reasonable amount of time.

\paragraph{Author's note}
We thank the reviewers for their
thoughtful comments.
The online appendix, the code used in this paper, the SAT instances,
and a collection of the certificates
are available at \url{uwaterloo.ca/mathcheck}---these
have also been archived at \url{doi.org/10.5281/zenodo.3842255}.

\bibliography{lams}

\end{document}


\title{A SAT-based Resolution of Lam's Problem: Online Appendix}
\author{}
\date{}
\maketitle
\addtocounter{table}{1}

\begin{figure}[h!]
\input{case1-partial.tikz}
\caption*{One partial completion of $A$ found
by our main search.  Black entries represent $1$,
white entries represent $0$, and each gray rectangle
contains a single $1$.
This particular completion consists of an
outside block along with
the inside blocks 3, 4, and~5.
As mentioned in the main text, the rows of the $A2$
have been reordered from lexicographic order---instead
the first row is incident to as many inside blocks as possible
and the number of mutually intersecting rows in $A2$
at the top is maximized.}
\end{figure}

\begin{theorem}\label{thm:record}
If the $A2$ SAT instances are solved with isomorphism removal performed
after the completion of each level then
the solver will record exactly one
representative from each equivalence class
of $A2$ completions.
\end{theorem}

\begin{proof}
Because isomorphism removal will be performed after
the final level (i.e., after finding a complete $A2$)
at most one solution from each equivalence class will be recorded.

Next, we must show that at least one solution from each equivalence
class will be recorded.
Suppose for the sake of contradiction that there is a solution $S$
that is not recorded (and no solutions isomorphic to $S$ are recorded).
Let $S_l$ denote the submatrix of $S$ up to and including level $l$.
Since $S_0$ (the empty matrix) will be recorded at the beginning of the
search and $S=S_6$ is not recorded there must be a level $l$ such that
$S_{l-1}$ is recorded but $S_{l}$ is not recorded.

Since $S_{l-1}$ is recorded the solver will examine all possible ways
of completing $S_{l-1}$ to level $l$ and thus will encounter $S_l$
in the search.  Since $S_l$ will be found but not recorded, there must be some isomorphic
$S'_l$ that was previously recorded.  Let $\varphi$ be the permutation
of rows and columns that sends $S_l$ to $S'_l$ and let $\varphi(S)$
denote applying the permutations of $\varphi$ to $S$
followed by resorting its rows
(as the lexicographic ordering of the rows after the first~$l$ levels may be disturbed).
This resorting does not affect the first~$l$ levels of $\varphi(S)$ as
these rows are already sorted and lexicographically greater than the rows
after the first~$l$ levels.
Thus $\varphi(S)$ is an $A2$ isomorphic to $S$ whose first $l$
levels consist of $S'_l$ and are therefore recorded.

Now there must be a new level $l'$ with $l<l'$ such that
the first $l'-1$ levels of $\varphi(S)$ are recorded but the first
$l'$ levels are not recorded.  Repeating the above reasoning
we find a recorded matrix isomorphic to the first $l'$ levels of $\varphi(S)$
and proceeding inductively we arrive at a recorded matrix containing all
levels and that is isomorphic to $S$---a contradiction.
\end{proof}

Note that one must be careful when isomorphism removal
is performed in order to ensure completeness---the proof of Theorem 1 requires that
isomorphism removal is performed following the completion
of each level.  The theorem will not necessarily hold if isomorphism removal
is performed following the completion of each row, for example.

\def\X{\phantom0}
\begin{table}\centering
\begin{tabular}{ccc}
Case  & Inequiv. & Total    \\
   \X1 & \hfill   20,129 & \hfill11,530,368 \\ 
   \X2 & \hfill   11,861 & \hfill9,014,784 \\ 
   \X3 & \hfill    2,501 & \hfill1,888,512 \\ 
   \X5 & \hfill    5,219 & \hfill7,704,576 \\ 
   \X6 & \hfill  111,538 & \hfill5,347,968 \\ 
   \X7 & \hfill   13,211 & \hfill 840,704 \\ 
   \X8 & \hfill  110,879 & \hfill3,542,944 \\ 
   \X9 & \hfill   87,807 & \hfill 702,456 \\ 
    11 & \hfill    7,059 & \hfill 225,632 \\ 
    12 & \hfill   10,635 & \hfill1,020,576 \\ 
    13 & \hfill   11,961 & \hfill 574,128 \\ 
    15 & \hfill   43,719 & \hfill 697,936 \\ 
    16 & \hfill    8,040 & \hfill 192,960 \\ 
    17 & \hfill    9,110 & \hfill 145,760 \\ 
    18 & \hfill    3,406 & \hfill 162,552 \\ 
    20 & \hfill   27,221 & \hfill3,918,672 \\ 
    21 & \hfill    9,410 & \hfill2,019,744 \\ 
    22 & \hfill   18,947 & \hfill 452,496 \\ 
    23 & \hfill    7,971 & \hfill 191,304 \\ 
    24 & \hfill   17,102 & \hfill 546,960 \\ 
    25 & \hfill   21,180 & \hfill 338,816 \\ 
    26 & \hfill   18,970 & \hfill 151,760 \\ 
    27 & \hfill    1,759 & \hfill  84,432 \\ 
    32 & \hfill    2,052 & \hfill 570,528 \\ 
    33 & \hfill   16,509 & \hfill 131,888 \\ 
    34 & \hfill      673 & \hfill  10,768 \\ 
    36 & \hfill    5,166 & \hfill 646,464 \\ 
    37 & \hfill    3,215 & \hfill  77,160 \\ 
    38 & \hfill    5,445 & \hfill 516,096 \\ 
    39 & \hfill      505 & \hfill  12,120 \\ 
    41 & \hfill   10,102 & \hfill  80,816 \\ 
    42 & \hfill    1,679 & \hfill  52,896 \\ 
    43 & \hfill    4,855 & \hfill  38,840 \\ 
    46 & \hfill       50 & \hfill  13,632 \\ 
    47 & \hfill      912 & \hfill  21,888 \\ 
    48 & \hfill      863 & \hfill  40,496 \\ 
    49 & \hfill      101 & \hfill1,255,680 \\ 
    50 & \hfill    1,064 & \hfill 291,600 \\ 
    51 & \hfill    3,685 & \hfill 174,384 \\ 
    52 & \hfill      604 & \hfill  43,488 \\ 
    53 & \hfill      182 & \hfill 103,104 \\ 
    54 & \hfill    1,458 & \hfill 409,536 \\ 
    55 & \hfill    3,656 & \hfill 115,224 \\ 
    56 & \hfill      794 & \hfill  12,432 \\ 
    57 & \hfill      601 & \hfill  56,400 \\ 
    58 & \hfill    4,188 & \hfill  33,320 \\ 
    60 & \hfill       90 & \hfill   6,264 \\ 
    63 & \hfill      196 & \hfill   6,000 \\ 
    64 & \hfill    1,135 & \hfill 101,760 \\ 
    65 & \hfill      787 & \hfill  35,948 \\ 
    66 & \hfill      168 & \hfill   2,648 \\ 
Total & \hfill  650,370 & \hfill56,157,420
\end{tabular}
\caption*{Counts for the number of inequivalent $A2$s (up to isomorphism)
and the total number of $A2$s in each case with no isomorphism removal
besides ordering the rows lexicographically.
Cases that are not listed have no $A2$s.}
\end{table}